\documentclass[american]{article}
\usepackage[T1]{fontenc}
\usepackage[utf8]{inputenc}
\usepackage{graphicx}

\makeatletter
\newcommand{\lyxaddress}[1]{
	\par {\raggedright #1
	\vspace{1.4em}
	\noindent\par}
}

\makeatother

\date{}

\usepackage{babel}
\begin{document}
\title{Non-equilibrium glassy arrest and discontinuous transitions at avoided
quantum critical points}
\author{J J Pulikkotil}
\maketitle

\lyxaddress{CSIR-National Physical Laboratory, Dr. K. S. Krishnan Marg, New Delhi
110012, India}
\begin{abstract}
The non-equilibrium dynamics of an order parameter confined by an
external constraint field are investigated within a spatially extended
Belitz-Kirkpatrick-Vojta framework. A generic non-monotonic dissipation
peak arises at avoided criticality due to the interplay between macroscopic
free-energy flattening and microscopic disorder-induced trapping.
Near this regime, suppressed deterministic forces enhance activated
trapping, leading to an effective violation of the fluctuation-dissipation
theorem and ultimate glassy arrest. At higher fields, the system undergoes
a discontinuous phase transition that bypasses this flat free-energy
region. These explicit analytical scalings establish a generic mechanism
for non-equilibrium arrest in disordered condensed matter systems.

Keywords : avoided quantum criticality, glassy dynamics, stochastic
thermodynamics, fluctuation-dissipation theorem, itinerant ferromagnets
\end{abstract}

\section{Introduction}

Classical models of non-equilibrium phase transitions often assume that increased external constraint demands proportionally higher dissipative costs \cite{Henkel2008}. Does this assumption hold during structural yielding or glassy freezing \cite{Liu1998, Debenedetti2001}? The interplay between macroscopic driving forces and microscopic disorder dictates the dynamics of random-field Ising models and activated scaling \cite{Nattermann1998, Fisher1986, Bouchaud1992}. To describe adaptation under constraint, the geometry of the underlying energy landscape must be evaluated alongside the stochastic dynamics of the system \cite{Imry1975, Stillinger1984}.

Itinerant ferromagnets such as LaCrGe$_3$ provide a physical realization of this competition, bypassing a continuous second-order critical point via a discontinuous first-order phase transition driven by quenched microscopic disorder \cite{Belitz1999, Taufour2016, Brando2016, Gati2021}. Recent first-principles evaluations emphasize that this pressure-driven phase boundary is intimately tied to the persistence of disordered local moments and spin fluctuations \cite{Himanshu2025, Himanshu2026}. By reframing the temporal evolution of an order parameter across a rugged energy landscape, the divergence of dissipative costs during avoided criticality can be analytically mapped. Within this framework, it is demonstrated that a generic non-monotonic dissipation peak arises precisely at avoided criticality due to the interplay between macroscopic free-energy flattening and microscopic disorder-induced trapping.

\section{Theoretical Framework}

The state of the target system is defined within a high-dimensional continuous order parameter vector space $\vec{\phi}(\mathbf{r}, t) = (\phi_1, \phi_2, \dots, \phi_N)$. If evolution under constraint is geometrically restrictive, local dynamics are projected onto an effective scalar field $\phi(\mathbf{r}, t) = |\vec{\phi}|_{constrained}$. Here, $\phi_{wt}$ represents the scalar magnitude of the stable equilibrium and $\phi = 0$ denotes the phase boundary.

Is the mean-field approximation sufficient to capture this avoided critical point? According to the Ginzburg criterion, critical fluctuations dominate near a continuous transition. However, by coupling the manifold to quenched disorder, the continuous transition is suppressed before the critical region is fully accessed. The effective dimensionality of the system becomes heavily constrained by local quenched disorder barriers, validating the scalar projection $\phi(\mathbf{r}, t)$.

The evolution of this effective scalar state is modeled on a spatially extended, sixth-order Belitz-Kirkpatrick-Vojta (BKV) Landau-Ginzburg manifold. A minimum of a sixth-order polynomial is required to support a first-order discontinuous transition while maintaining global stability \cite{Belitz1999}. The macroscopic free energy functional, $\mathcal{F}_{LG}[\phi]$, is defined as an integral over the spatial volume $V$:

\begin{equation}
\mathcal{F}_{LG}[\phi] = \int_V d\mathbf{r} \left[ \frac{1}{2} \kappa (\nabla \phi)^2 + \frac{1}{2} a(R_A) \phi^2 - \frac{1}{4} \beta \phi^4 + \frac{1}{6} \gamma \phi^6 - \lambda R_A \phi \right]
\end{equation}

Here, $\kappa$ governs the energy penalty for spatial variations. $R_A$ represents the external constraint field intensity. The mass-tuning parameter $a(R_A) = T_0 + \alpha R_A$ determines primary curvature, where $T_0$ defines the intrinsic thermodynamic stability of the unconstrained state. The term $\lambda R_A \phi$ acts as the macroscopic symmetry-breaking field.

To formalize the quenched microscopic disorder inherent to the BKV manifold, a high-frequency ruggedness potential is introduced: $\mathcal{U}_{rugged}(\phi) = -K \sin^2(k_d \phi)$. Here, $k_d = \pi/\xi_d$ establishes the characteristic inverse length scale ($\xi_d$) of the coarse-grained quenched disorder. While this periodic potential serves as a coarse-grained representation, equivalent behavior is obtained for random disorder realizations, capturing the generic physics of activated trapping analogous to the Bouchaud trap model \cite{Bouchaud1992}. The total landscape is $\mathcal{F}_{total}[\phi] = \mathcal{F}_{LG}[\phi] + \mathcal{U}_{rugged}(\phi)$.

This dual-scale formulation characterizes phase jumps through dimensionless parameterization, defining the constraint field as $r = R_A / R_{A, crit}$. In the linear response limit ($r \ll 1$), the system exhibits stable thermal fluctuations. At the degenerate limit ($r \to 1$), the macroscopic gradient flattens, trapping the system within the effective scalar coordinate via the quenched ruggedness $K$. In the discontinuous jump limit ($r \gg 1$), the external field overwhelms internal structural parameters, forcing an instantaneous transition across $\phi = 0$.

\section{Stochastic Thermodynamics and Arrest}

Structural evolution is governed by the time-dependent Ginzburg-Landau equation \cite{Hohenberg1977, Kramers1940}:

\begin{equation}
\frac{\partial \phi(\mathbf{r}, t)}{\partial t} = - \Gamma \frac{\delta \mathcal{F}_{total}}{\delta \phi(\mathbf{r}, t)} + \sqrt{2 \Gamma T_{eff}(t)} \eta(\mathbf{r}, t)
\end{equation}

Here, $\eta(\mathbf{r},t)$ is spatiotemporal Gaussian white noise. Applying Markovian white noise alongside a dynamically depleting dissipative bath $R_B(t)$ necessitates an adiabatic decoupling. It is assumed that local thermal fluctuations relax infinitely faster than the macroscopic depletion rate ($\tau_{fluct} \ll \tau_{depletion}$), allowing the noise to remain uncorrelated while the global energy reservoir drains.

How is the thermodynamic cost of this evolution quantified? A finite energy reservoir, $R_B(t)$, acts as a dissipative bath governed by a generalized entropy production metric \cite{Kanazawa2012, Seifert2012}:

\begin{equation}
\dot{\Sigma}_B = \int_V d\mathbf{r} \left[ \mu_B (\nabla \phi)^2 + \mu_{\phi} (\phi(\mathbf{r}, t) - \phi_{wt})^2 \right] + \Sigma_{basal}
\end{equation}

This integral represents a minimal quadratic approximation consistent with coarse-grained entropy production in stochastic thermodynamics. The spatial gradient penalty, $\mu_B (\nabla \phi)^2$, ensures the reservoir measures the energetic cost of maintaining frustrated interfaces against quenched traps. Stochastic mobility is modeled by coupling the effective kinetic temperature to this depleting budget: $T_{eff}(t) = T_{strat} (R_B(t)/R_{B,0})$. This serves as a phenomenological description of mobility reduction under dissipation \cite{Cugliandolo1997, Berthier2011}.

As the effective temperature $T_{eff}(t)$ decays, equilibrium statistical mechanics fails to adequately describe the phase stability. The system exhibits an effective violation of the fluctuation-dissipation theorem (FDT). Specifically, the two-time correlation function $C(t, t') = \langle \phi(t) \phi(t') \rangle$ and the corresponding response function $R(t, t')$ diverge from the standard equilibrium proportionality $T_{strat} R(t, t') = \partial_{t'} C(t, t')$. Instead, the local kinetics are governed by an extended fluctuation-dissipation ratio defined by the slowly decaying $T_{eff}(t)$ \cite{Zwanzig2001}. Sustained spatial strain energy strictly drains $R_B$, domain walls freeze against the epistatic topology, and the system undergoes non-equilibrium glassy arrest \cite{Anderson1958, Binder1986}.

\section{Numerical Implementation and Parameterization}

To evaluate the temporal evolution of the order parameter and the corresponding thermodynamic costs, the time-dependent Ginzburg-Landau equation (Equation 2) was solved numerically. The stochastic differential equation was integrated utilizing a standard Euler-Maruyama scheme to appropriately handle the additive spatiotemporal Gaussian white noise $\eta(\mathbf{r},t)$. The integration was performed over a discretized spatial grid representing the effective scalar field, with periodic boundary conditions applied to minimize finite-size edge effects.

The integration timestep $\Delta t$ was selected to be significantly smaller than the characteristic relaxation time of the unconstrained system to ensure numerical stability and to properly capture the high-frequency dynamics of the ruggedness potential $\mathcal{U}_{rugged}(\phi)$. The macroscopic depletion rate of the dissipative bath $R_B(t)$ was updated dynamically at each macroscopic time step according to Equation 3, feeding back into the instantaneous effective temperature $T_{eff}(t)$. This approach strictly enforces the adiabatic decoupling assumption, allowing local fluctuations to equilibrate against the instantaneous landscape geometry.

\begin{figure}
\includegraphics[scale=0.25]{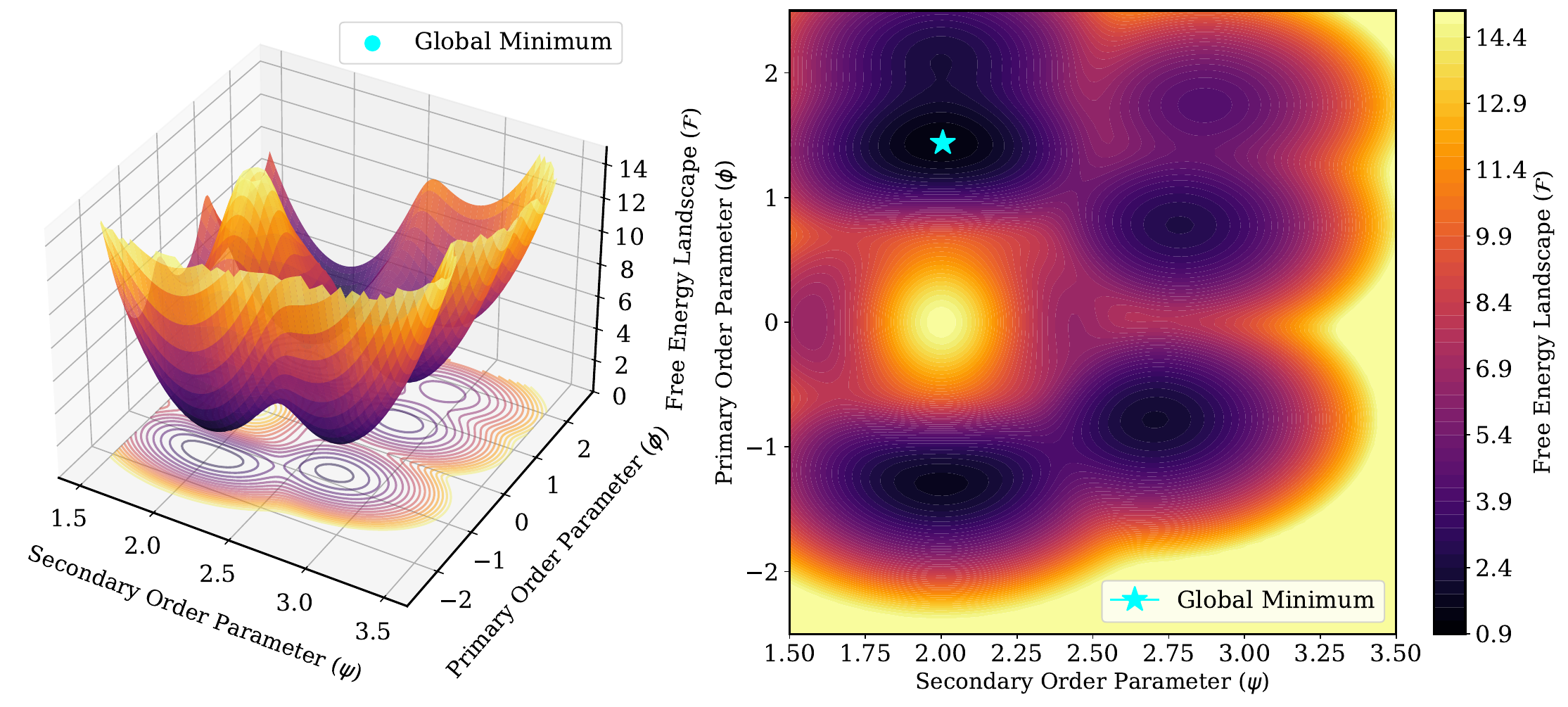}
\includegraphics[scale=0.25]{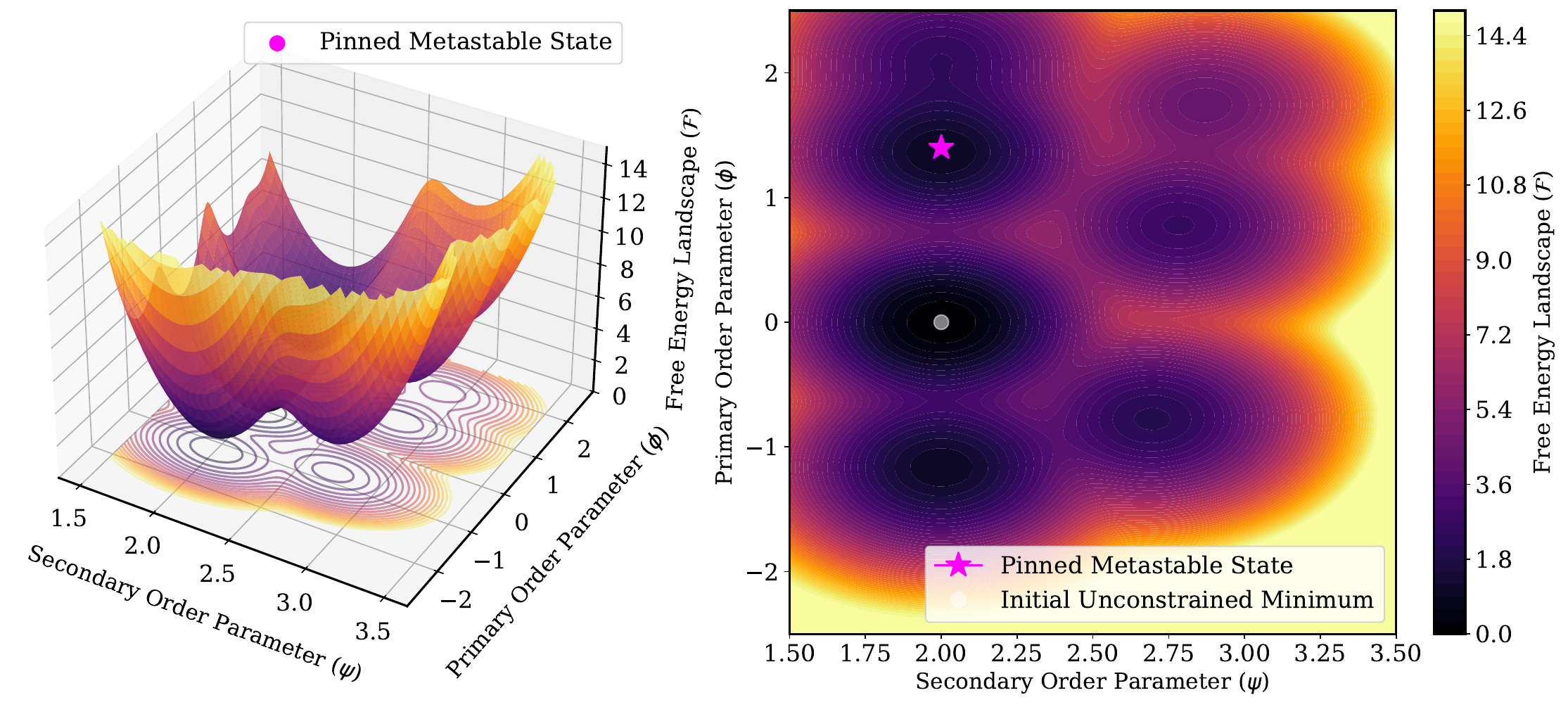}
\caption{Phenomenological visualization of the constraint-induced free energy landscape $\mathcal{F}$ across the primary ($\phi$) and secondary ($\psi$) order parameters. (a) Top panel, Strong Field Regime ($R_A = 15$): Under an intense external constraint field, the macroscopic gradient forces the order parameter trajectory toward an orthogonal minimum. (b) Bottom panel, Zero Field Regime ($R_A = 0$): Upon removal of the external field, emergent high-frequency quenched barriers block the return trajectory. Lacking sufficient thermal activation ($T_{eff}$) to overcome these microscopic traps, the system undergoes non-equilibrium glassy arrest.}
\label{fig:1}
\end{figure}

The mechanism is visualized through the deformation of the configuration space (Figure \ref{fig:1}). An external constraint field destabilizes the equilibrium state, directing the gradient orthogonally. Upon removal of the field, the initial state is blocked by emergent barriers. Lacking sufficient thermal fluctuations, the system falls into a pinned state, marking the onset of disorder-induced metastability.

\begin{figure}
\includegraphics[scale=0.25]{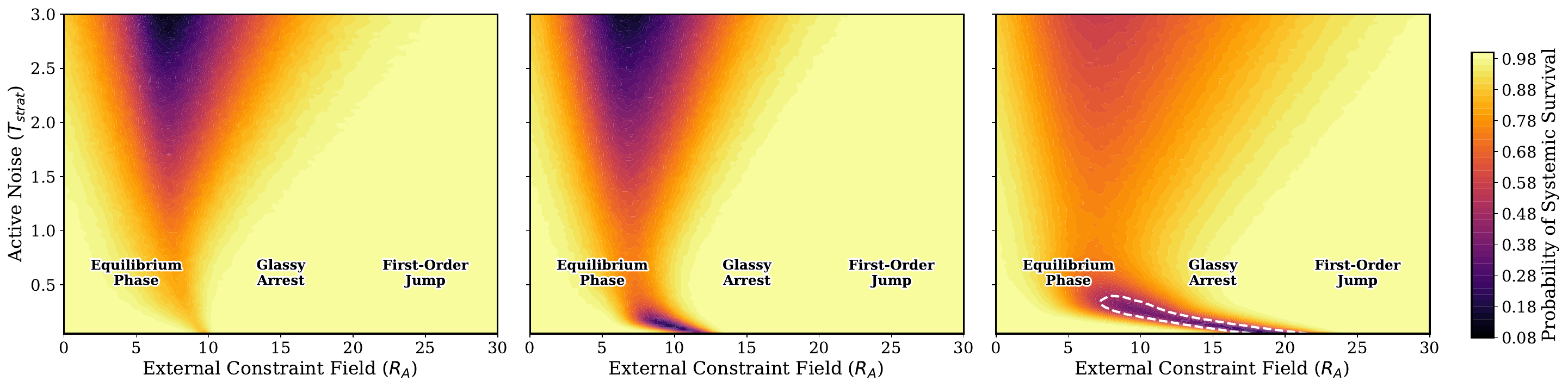}
\caption{Persistence probability phase diagrams across active noise ($T_{strat}$) and external constraint field intensity ($R_A$). A U-shaped region of phase instability forms near the critical threshold. This trap deepens and widens significantly as quenched microscopic ruggedness ($K$) increases, dominating the macroscopic free-energy flattening and increasing the density of metastable states (configurational entropy).}
\label{fig:2}
\end{figure}

Figure \ref{fig:2} maps the phase persistence probability. At low constraint intensity ($r \ll 1$), the equilibrium phase remains stable. As the constraint approaches the avoided critical threshold ($r \to 1$), persistence plummets, establishing a glassy arrest zone. In the smooth theoretical limit ($K = 0.0$), failure results from macroscopic softening. With quenched traps ($K = 1.5$ to $4.0$), local minima capture physical domains lacking adequate stochastic mobility, widening the zone of disorder-induced metastability.

The driver of this glassy state is an increase in configurational entropy. As the deterministic macroscopic gradient flattens near the avoided critical point, the accessible metastable microstates generated by $K$ diverge. Beyond the critical threshold ($r \gg 1$), the extreme gradient resolves this metastability, and the system yields via a discontinuous jump.

\section{Scaling Limit of Dissipation}

\begin{figure}
\includegraphics[scale=0.25]{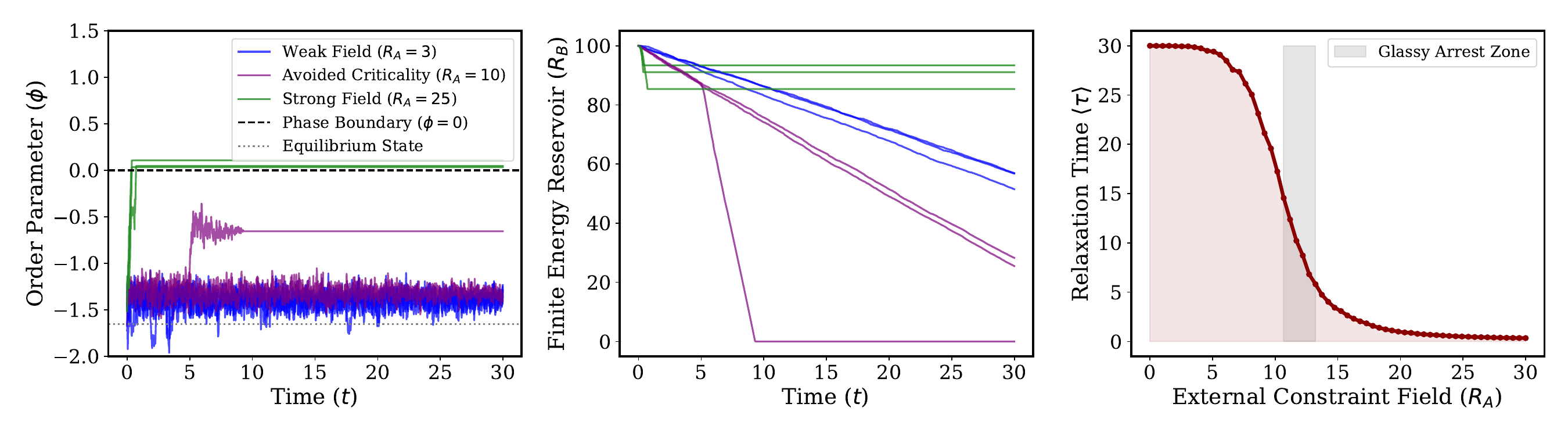}
\caption{Non-equilibrium kinetic trajectories of the order parameter field. \textbf{Left panel:} First-order discontinuous trajectories of the effective scalar field under varying constraint fields. \textbf{Middle panel:} Corresponding strain energy and resource depletion of the finite dissipative bath ($R_B$). \textbf{Right panel:} Divergence of relaxation time precisely at the avoided critical threshold ($R_A \approx 10$). This critical slowing down is driven by the pinning of the spatial correlation length against localized quenched traps.}
\label{fig:3}
\end{figure}

Kinetic trajectories (Figure \ref{fig:3}) clarify this scaling behavior. At $r = 1$, glassy arrest leads to rapid depletion of the finite reservoir. As the spatial correlation length $\xi$ attempts to diverge, the true relaxation time is strictly controlled by activated dynamic scaling \cite{Fisher1986}. Because $T_{eff}$ decays as the reservoir drains due to accumulated entropy production, the relaxation time formally diverges. This divergence acts as an early-warning signal for the critical transition \cite{Scheffer2009}.

\begin{figure}
\includegraphics[scale=0.25]{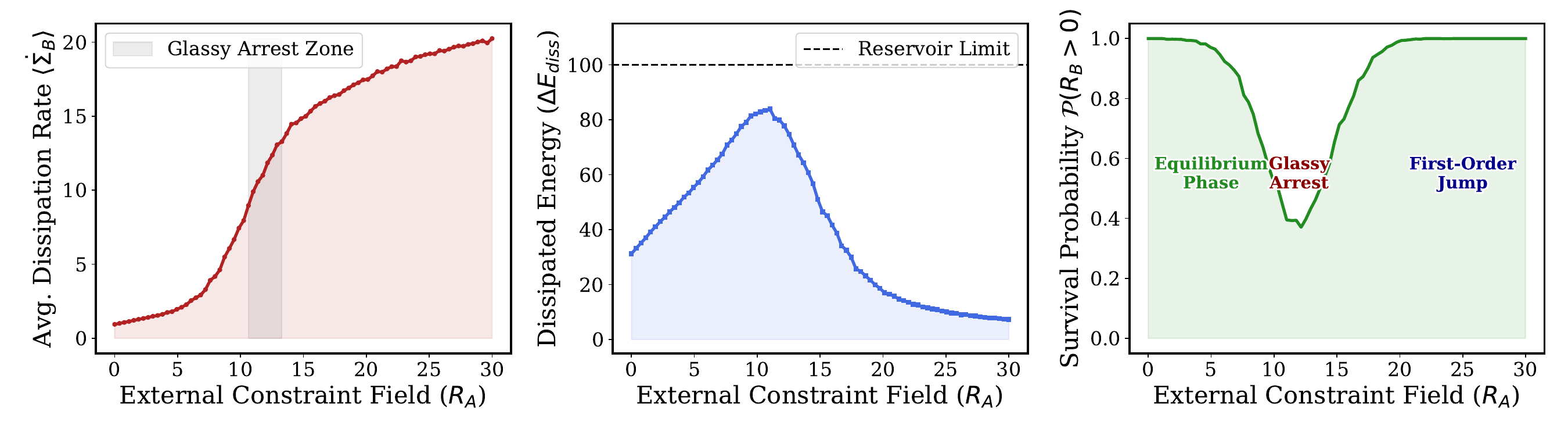}
\caption{Thermodynamic signature and phase stability. \textbf{Left panel:} Divergence of the average entropy production rate ($\langle \dot{\Sigma}_B \rangle$). \textbf{Middle panel:} Total dissipated energy ($\Delta E_{diss}$). The cost peaks non-monotonically precisely during the glassy arrest regime, while the first-order discontinuous jump ($R_A > 20$) bypasses the flat free-energy region entirely. \textbf{Right panel:} Persistence probability mapping the characteristic U-shaped region of phase instability caused by disorder-induced metastability.}
\label{fig:4}
\end{figure}

Figure \ref{fig:4} quantifies the total dissipated energy $\Delta E_{diss}$. When evaluated exactly at the avoided critical limit ($r \to 1$), the macroscopic deterministic force vanishes. The dynamics are entirely dictated by the variance of the quenched disorder, parameterized by $K^2$. This scaling behavior can be formally derived by relating activated escape times to localized barrier heights. The activated escape time $\tau_{esc}$ from a metastable state is defined by the Arrhenius relation:

\begin{equation}
\tau_{esc} = \tau_0 \exp\left( \frac{\Delta E}{T_{eff}} \right)
\end{equation}

where the characteristic barrier height scales with the ruggedness variance, $\Delta E \propto K^2$. Since entropy production accumulates locally over this residence period, the total dissipated energy prior to thermal escape is the integral of the production rate:

\begin{equation}
\Delta E_{diss} = \int_0^{\tau_{esc}} \dot{\Sigma}_B dt \approx \langle \dot{\Sigma}_B \rangle \tau_{esc}
\end{equation}

By observing that the generalized entropy production rate is proportional to the inverse of the initial kinetic mobility, $\langle \dot{\Sigma}_B \rangle \propto (\Gamma T_{strat})^{-1}$, substituting the Arrhenius residence time yields the formal scaling relation for the non-monotonic dissipation maximum:

\begin{equation}
\max(\Delta E_{diss}) \propto \frac{K^2}{\Gamma T_{strat}}
\end{equation}

Crucially, this scaling behavior holds strictly in the regime where $T_{eff}$ controls activated dynamics. Outside this regime, such as at limits where kinetic energy overwhelms the ruggedness potential, or at the absolute zero limit where thermal activation is completely suppressed, the Arrhenius relation breaks down. However, within the glassy arrest zone, this analytical limit indicates that maximum dissipation is dominated by disorder-induced metastability rather than the macroscopic gradient.

Conversely, the high-field phase jump ($r \gg 1$) is thermodynamically efficient, as the instantaneous annihilation of the phase boundary bypasses the flat free-energy region entirely. This mechanism is robust to dimensionality and system size, persisting generically wherever a flattening continuous gradient competes with quenched microscopic ruggedness.

\section{Conclusion}

Phase stability is determined by the width of the flat free-energy region, the system's susceptibility to disorder-induced trapping rather than its structural capacity to resist a continuous external field. In itinerant ferromagnets, predicted glassy arrest should manifest as non-Gaussian tails and crackling noise in Barkhausen spectroscopy, indicating avalanche dynamics as domain walls struggle against pinning sites \cite{Sethna2001, Vojta1999}.

Investigating the strict zero-temperature limit could determine if macroscopic quantum tunneling provides an alternative probabilistic pathway through the phase boundary \cite{Caldeira1981}. Future formulations must also incorporate annealed disorder, where quenched traps evolve in response to the order parameter. Coupling $\phi(\mathbf{r}, t)$ to a secondary, slowly relaxing field would enable the investigation of aging, rejuvenation, and hysteresis, characteristic of dynamic heterogeneity \cite{Berthier2011}. The thermodynamic singularity observed at the avoided critical point establishes that external fields leading to systemic failure are those that successfully flatten the macroscopic gradient, inducing non-equilibrium glassy arrest against quenched microscopic disorder.

\section*{Declarations}

\textbf{Conflicts of Interest:} The author declares no relevant financial or non-financial competing interests that could have influenced the outcomes, interpretations, or conclusions presented in this work.

\textbf{Use of Artificial Intelligence (AI) Tools:} The author declares the use of Generative AI exclusively for the purpose of linguistic refinement. The author has thoroughly reviewed and verified all content and assumes full responsibility for the integrity and accuracy of the finalized manuscript.

\textbf{Data Availability:} All theoretical derivations are contained within the manuscript. The computational codes and numerical data utilized to generate the kinetic trajectories and thermodynamic scalings are available from the corresponding author upon reasonable request.

\end{document}